\documentclass[conference]{IEEEtran}
\IEEEoverridecommandlockouts

\usepackage{cite}
\usepackage{amsmath,amssymb,amsfonts}
\usepackage{algorithmic}
\usepackage{graphicx}
\usepackage{textcomp}
\usepackage{xcolor}
\usepackage{hyperref}
\usepackage{siunitx}
\usepackage{hyperref}
\usepackage{lipsum}%
\usepackage[pscoord]{eso-pic}
\newcommand{\placetextbox}[3]{%
  \setbox0=\hbox{#3}
  \AddToShipoutPictureFG*{
    \put(\LenToUnit{#1\paperwidth},\LenToUnit{#2\paperheight}){\vtop{{\null}\makebox[0pt][c]{#3}}}%
  }%
}%

\usepackage{siunitx}

\def\BibTeX{{\rm B\kern-.05em{\sc i\kern-.025em b}\kern-.08em
    T\kern-.1667em\lower.7ex\hbox{E}\kern-.125emX}}
\begin{document}

\title{Compression of executable QR codes or sQRy for Industry: an example for Wi-Fi access points\vspace{-0.2cm}\\
\thanks{This work has been partially funded by CNR under the project ``Executable QR codes (EQR) - GORU IEIIT'' (DIT.AD001.212).}
}

\author{
    \IEEEauthorblockN{
    Stefano Scanzio\IEEEauthorrefmark{1},
    Gabriele Formis\IEEEauthorrefmark{1}\IEEEauthorrefmark{2},
    Pietro Chiavassa\IEEEauthorrefmark{1},
    Lukasz Wisniewski\IEEEauthorrefmark{3},
    Gianluca Cena\IEEEauthorrefmark{1}
    }
    
    \IEEEauthorblockA{\IEEEauthorrefmark{1}National Research Council of Italy (CNR--IEIIT), Italy. \IEEEauthorrefmark{2}Politecnico di Torino, Italy.}    
    \IEEEauthorblockA{\IEEEauthorrefmark{3}Institute Industrial IT - inIT, Technische Hochschule OWL, Germany.}
    Emails: \{stefano.scanzio, gabrieleformis, pietrochiavassa, gianluca.cena\}@cnr.it, lukasz.wisniewski@th-owl.de\vspace{-0.3cm}
    }

\placetextbox{0.5}{1}{This is the author's version of an article that has been published.}
\placetextbox{0.5}{0.985}{Changes were made to this version by the publisher prior to publication.}
\placetextbox{0.5}{0.97}{The final version of record is available at \href{https://doi.org/10.1109/WFCS63373.2025.11077642}{https://doi.org/10.1109/WFCS63373.2025.11077642}}%
\placetextbox{0.5}{0.05}{Copyright (c) 2025 IEEE. Personal use is permitted.}
\placetextbox{0.5}{0.035}{For any other purposes, permission must be obtained from the IEEE by emailing pubs-permissions@ieee.org.}%

\maketitle

\begin{abstract}
Executable QR codes, or sQRy, is a technology dated 2022 that permits to include a runnable program inside a QR code, enabling interaction with the user even in the absence of an Internet connection. sQRy are enablers for different practical applications, including network equipment configuration, diagnostics, and enhanced smart manuals in industrial contexts. Many other non-industry-related fields can also benefit from this technology.

Regardless of where sQRy are used, text strings are among the most commonly embedded data. However, due to strict limitations on the available payload, the occupancy of strings limits the length of the programs that can be embedded. In this work, we propose a simple yet effective strategy that can reduce the space taken by strings, 
hence broadening sQRy applicability.
\end{abstract}

\section{Introduction}
QR codes are a consolidated technology invented by Denso Wave Automotive in 1994 and standardized in 2015~\cite{STD-QRcode}.
They are exploited in many applicative scenarios including 
smart documents~\cite{10594158}, digital payment~\cite{10698108}, education~\cite{10619878},
traceability, quality insurance~\cite{15}, and security~\cite{IoT1}. An in-depth literature survey about QR codes can be found in~\cite{10492995}.
In their common usage pattern, QR codes encode a uniform resource identifier (URI) that points to a remote application running on a web server and interacting with the user.
In case of limited, intermittent, or absent Internet connection, this is not possible because the server cannot be reached.

To overcome this limitation, executable QR codes (eQR codes), which are also known as sQRy (\url{https://www.sqry.org}), were presented in 2022~\cite{9921530}. 
Since the executable code is included in their payload, no Internet access is required to interact with them.
This characteristic makes sQRy a valuable technology in all those areas characterized by poor geographic connectivity. 
In industrial settings, there are remote installations such as s, but also those in desert areas (e.g., wind power plants) or in the mountains (e.g., hydroelectric plants or remote sensors connections) where Internet connection is not available. 
In all these contexts, sQRy-assisted interaction may help in configuring local network equipment, e.g., (industrial) Ethernet, Wi-Fi, and fiedbuses, diagnosing potential failures and suggesting recovery actions for both network devices and the physical and software components of production machinery.

The main problem of QR codes is their limited capacity in terms of payload, which in its most spacious configuration (i.e., $177\times 177$ matrix of version 40 and with low compression level) can store up to $\SI{2953}{bytes}$. 
In common applications of sQRy, strings constitute a significant portion of the program. For instance, in the example presented in this paper, they occupy $91.6\%$ of the entire compiled program.

The main goal of this work is to define a simple yet effective compression method for strings. 
All details about its complete integration into the sQRy technology are presented and finalized.
Importantly, it has been conceived to be extensible:
due to the way it is implemented, it does not prevent the adoption of new, more efficient compression algorithms at a later time.

In the next section of this work, the sQRy technology and some information about the programs that can be included are illustrated. Section~\ref{sec:integration} provides details about the string compression method and its integration. These concepts are applied to a concrete example concerning network equipment in Section~\ref{sec:example}, just before the concluding remarks.

\section{sQRy and QRtree dialect}
\begin{figure}[b]
\vspace{-0.5cm}
\begin{center}
\includegraphics[width=0.65\linewidth]{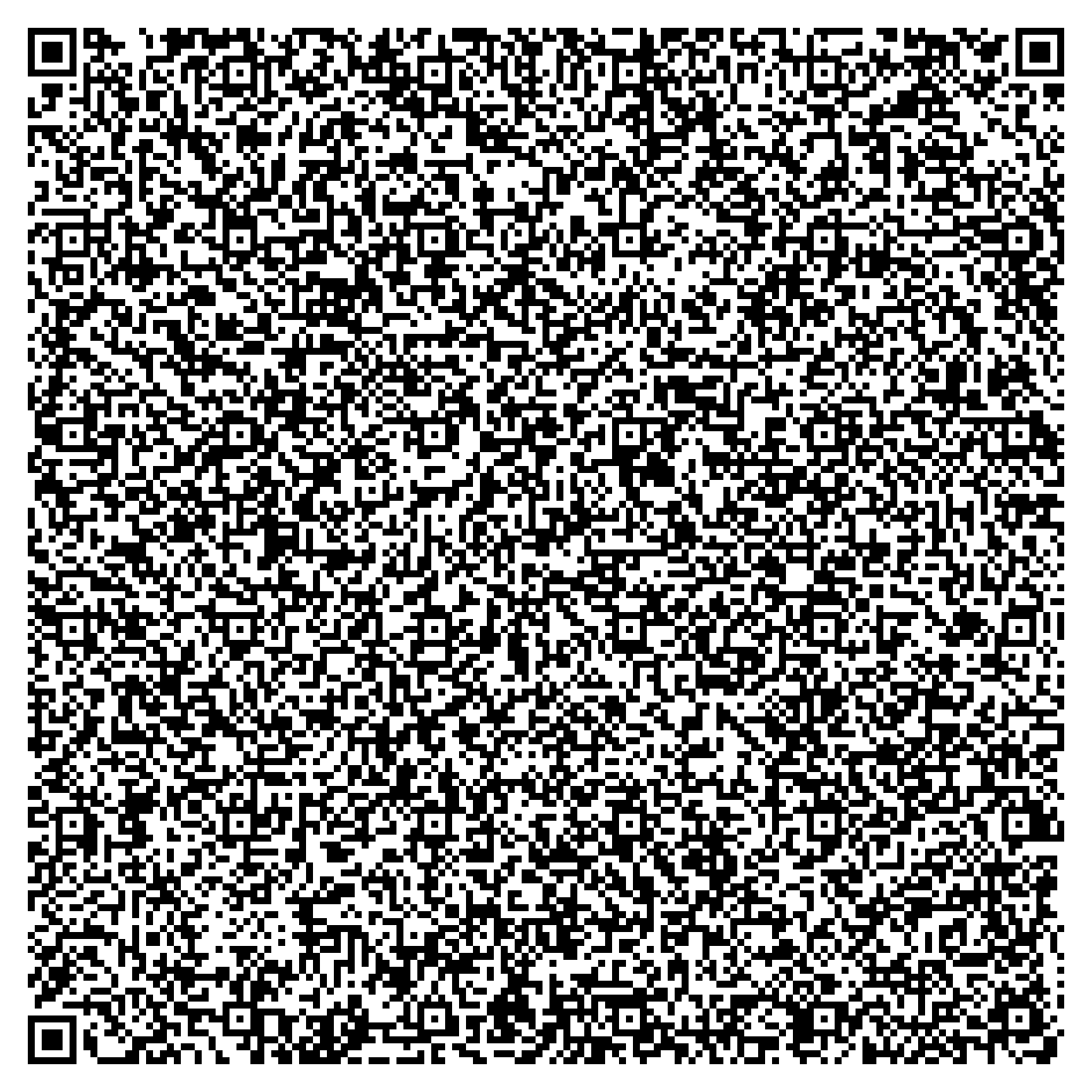}
\end{center}
\vspace{-0.5cm}
\caption{sQRy of the program reported in Fig.~\ref{fig:high_level}.}
\label{fig:sQRy}
\end{figure}

A sQRy looks like a conventional QR code as the one reported in Fig.~\ref{fig:sQRy}, which contains a program to manage the Wi-Fi equipment described in Section~\ref{sec:example}. 
\begin{figure}[b]
\vspace{-0.5cm}
\begin{center}
\includegraphics[width=1.0\linewidth]{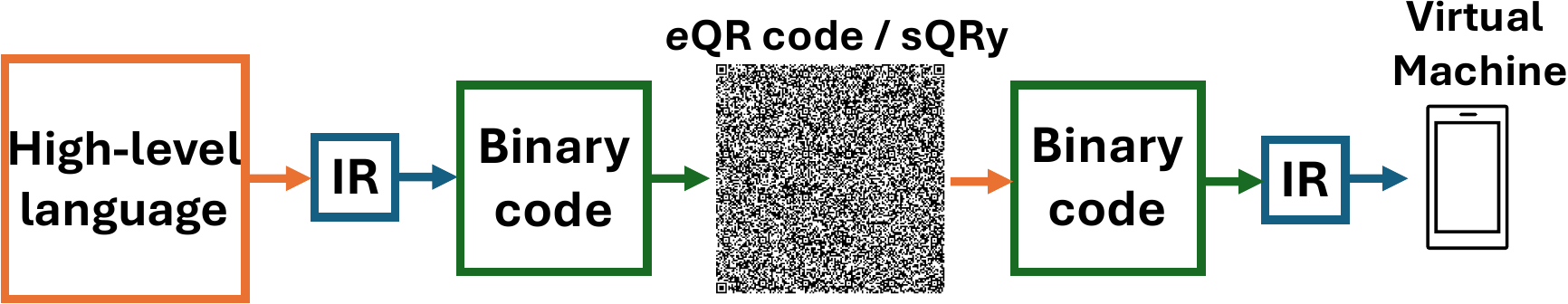}
\end{center}
\vspace{-0.3cm}
\caption{Generation and execution chain of sQRy.}
\label{fig:chain}
\end{figure}
sQRy cannot be natively interpreted by commercial smartphones and QR code readers, but specific software chains have to be defined to generate and execute them, as reported in Fig.~\ref{fig:chain}. 
In the \textit{generation} chain, starting from a high-level programming language (e.g., textual or graphical), an intermediate representation (IR) composed of simple instructions (e.g., three-address code) is obtained, which is then translated into a binary representation that can be easily used to generate a QR code using open-source libraries. 
Instead, the \textit{execution} chain concerns the execution of the embedded program. 
Specific libraries are used to acquire and decode the QR code picture and get back the binary representation, which is then translated into an IR that is run through a specifically programmed virtual machine in the end device (e.g., a smartphone).

The most important part of these chains is the definition of the IR and how it translates into a binary representation and vice versa. 
In fact, the main requirement of IR is compactness, for tackling the limitations on the maximum payload of QR codes, which is why several IRs (that are named dialects in the sQRy terminology) were defined.

This work is based on QRtree, which
is the dialect that reached the highest level of maturity and is aimed to encode decision trees. 
It was defined in~\cite{10492995} and in the specification documents \cite{QRscript-spec, QRtree-spec},
and shows a very good encoding efficiency.
The second dialect defined for sQRy is QRind~\cite{10710739}, which was conceived for industrial applications and permits to embed simple machine learning algorithms.

\section{Compression method and sQRy integration}
\label{sec:integration}
The portion of space occupied by strings in typical sQRy applications can be quite large. 
For instance, in the example provided in the next Section~\ref{sec:example}, $91.6\%$ of the payload is used to store strings. 
Therefore, the integration of techniques aimed at reducing string occupancy in QRtree was investigated in this work.
To this end, the integration mechanism was divided into two sub-activities: proposing a simple compression algorithm and integrating the capability of compressing strings within the QRtree dialect.
It is important to note that, as with many other compression algorithms, the process used to integrate compressed strings in sQRy is decoupled from the heuristics used to drive compression. This separation is intentional and permits the future standardization of the method without limiting improvements in the compression technique. For this reason, the presentation of only one simple compression method is not a limitation but just a first proposal, which will be improved in the future.

\subsection{Compression method}
We propose a very simple and effective compression method that is based on the collection of statistics about words composing the strings. As previously introduced, the selection of a simple compression technique is motivated by the focus of this work, which is on the integration of the technique on QRtree, rather than on compression performance.
Within each string, we identify as words the following regular expression \texttt{[a-zA-Z0-9\_\textbackslash.\textbackslash-]+}, i.e., any combination of characters, numbers, and symbols \texttt{\_} (underscore), \texttt{.} (dot), and \texttt{-} (minus). 
All the other characters (i.e., the complement \texttt{[\textasciicircum a-zA-Z0-9\_\textbackslash.\textbackslash-]+} of the previous set) are considered as separators between words. It is important to notice that slightly altering the definition of a word does not significantly impact the results in terms of compression efficiency.

\vspace{0.2cm}
Given the following three sentences (the same as in Fig.~\ref{fig:coding}.a):
\begingroup
\footnotesize{}
\begin{verbatim}
 "Wi-Fi activity detected"
 "Wi-Fi activity not detected"
 "Wi-Fi 802.11ax activity at 9600 Mbps"
\end{verbatim}
\endgroup
the proposed method selects the words repeated at least two times with a length (in terms of number of characters) greater than $2$. 
Three characters is the minimum word size that permits compression when the integration in the sQRy is considered, due to the additional bits needed to manage the concatenation of words in the dictionary and sub-strings. 
The algorithm generates a simple dictionary composed of the words \texttt{"Wi-Fi"}, \texttt{"activity"}, and \texttt{"detected"} with $3$, $3$, and $2$ occurrences, respectively.

The simple idea behind this compression method is to include the dictionary at the beginning of the sQRy and assign each word in the dictionary a unique binary identifier that can be used in the place of that word. 
The main contribution of this preliminary work is a detailed description of how this can be integrated into sQRy.

\subsection{sQRy integration}
\label{sub:sQRy_integration}
This section details the integration of string compression in the QRtree dialect, which is the only one completely defined at the moment of writing. 
Nevertheless, the proposed solution can be generalized to other dialects and thus to all types of sQRy, because strings are widely used in all application contexts. The proposed improvement is not backward compatible with the current version of QRtree. However, the \textit{version} field~\cite{QRscript-spec} could be used to identify a new version of the dialect, which can integrate the proposed technique.
Strings can be coded in \textit{ASCII-7} to have a more compact notation, or in \textit{UTF-8} to deal with a variety of alphabets, for modern and old languages, and symbolic and special characters such as mathematical symbols and emoji. Other encodings, such as \textit{ISO-8859}, were excluded because the set of characters they can encode is a subset of those encodable in \textit{UTF-8}. \textit{UTF-16} and \textit{UTF-32} were not considered because they are less efficient and do not offer any tangible benefit if compared with \textit{UTF-8}.

\begin{figure*}[t]
\begin{center}
\includegraphics[width=0.90\linewidth]{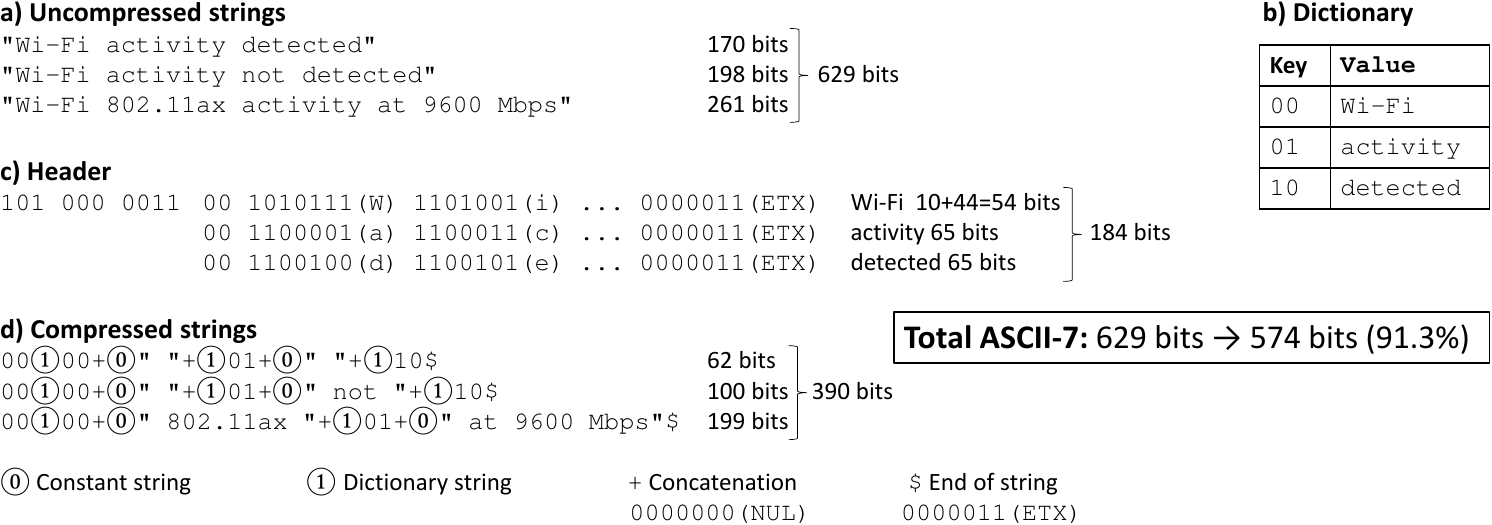}
\end{center}
\vspace{-0.5cm}
\caption{Encoding of header and strings to improve compression (in the example all the strings are coded in \textit{ASCII-7}).}
\label{fig:coding}
\vspace{-0.5cm}
\end{figure*}

The running example in Fig.~\ref{fig:coding} shows how the compression technique is incorporated into a sQRy. 
Fig.~\ref{fig:coding}.a reports three strings that, without compression, would occupy \SI{629}{bits} overall for the \textit{ASCII-7} encoding. Similar considerations can be made for \textit{UTF-8}. 

The dictionary (Fig.~\ref{fig:coding}.b) can be easily stored into the header of the QRtree dialect, because this option was already defined in the relevant specification document~\cite{QRtree-spec}.

With reference to Fig.~\ref{fig:coding}.c, the integration can be done with the specific \texttt{DICT\_LOCAL} command, identified by the sequence of bits \texttt{101} and followed by the constant sequence \texttt{000}, as expected by the initial definition of dictionaries in QRtree~\cite{QRtree-spec}. 
Finally, an integer number coded with exponential coding notation for signed numbers on $4$ bits (\texttt{0011}) represents the number of words stored in the dictionary, followed by the coding of the words, each terminated with an \texttt{ETX} character (\texttt{0000011} in \textit{ASCII-7} and \texttt{00000011} in \textit{UTF-8}). 
The exponential coding~\cite{QRtree-spec} allows for efficient encoding of small integers while not having an upper limit on the size of integers that can be encoded.
The first two bits of each word/string are the coding: \texttt{00} for \textit{ASCII-7} and \texttt{01} for \textit{UTF-8}. In the example of the figure, the dictionary contains three words with an associated key encoded on a number of bits equal to $n_{\mathrm{bits}} = \lceil \log_2(n_{\mathrm{words})} \rceil$ (in this case $\lceil \log_2 (3)\rceil=2$).

Regarding strings, instead, the way they are encoded has been completely redefined compared to~\cite{QRtree-spec}. 
The technique proposed in this work extends the definition of strings, permitting the concatenation between any number of \textit{constant} and \textit{dictionary} sub-strings. 
Each concatenated string, which is composed of one or more sub-strings, starts with two bits that identify its coding (\textit{ASCII-7} or \textit{UTF-8}).
Each sub-string 
starts with one bit to distinguish between a constant sub-string (\texttt{0}) and a dictionary sub-string  (\texttt{1}). 
In the case of a constant sub-string, the \texttt{0} bit is followed by the coding of the string (in either \textit{ASCII-7} or \textit{UTF-8}). 
Instead, in the case of a dictionary sub-string, the \texttt{1} bit is followed by the key of the related word, as determined by the order in which words are listed in the dictionary defined in the header. 
To separate two adjacent words the \textit{null} character (NUL), which is coded as \texttt{0000000} in \textit{ASCII-7} and \texttt{00000000} in \textit{UTF-8}, is used (see character \texttt{+} in Fig.~\ref{fig:coding}.d). 
Instead, the termination of a string is coded with the \textit{end of string} character (ETX), which corresponds to \texttt{0000011} in \textit{ASCII-7} and  \texttt{00000011} in \textit{UTF-8} (see character \texttt{\$} in Fig.~\ref{fig:coding}.d).

In the simple example of the figure, the proposed compression method permits lowering the total occupancy of the three strings from \SI{629}{bits} (original size) to \SI{574}{bits} ($91.3\%$ of the uncompressed counterpart).
Space saving is not very high, as compression efficiency depends on the specific selection of the dictionary entries.
In the case of the example, it is also limited because the number of repetitions of the dictionary words in the original strings to be compressed is $3$ at best.

\section{Example}
\label{sec:example}
To analyze the proposed compression method, an example of sQRy related to the configuration and usage of a \mbox{Wi-Fi} access point is presented and discussed. 
This example may be of interest in industrial scenarios for several reasons: 
a)~in some cases, accessing the Internet within an industrial plant is not easy, due, for instance, to its position, as for petrochemical industries; b) access to the local plant network can be forbidden for security reasons; and, c) the device that needs to be configured is the one that permits accessing the network, as in the case of APs.

\begin{figure}[t]
\fontsize{7}{8}\selectfont
\linespread{0.85}
\begin{verbatim}
input "Operation?"

if "Check status":
   input "What led?"
   if "Power":
      input "What color?"
      if "Amber initial":
         print "Starting/Getting IP address" exit
      else if "Green":
         print "Operating standalone mode" exit
      else if "Blue":
         print "Operating insight mode" exit
      else if "Blinking Amber":
         print "Firmware update" exit
      else if "Blinking multicolor":
         print "Mesh setup in progress" exit
      else if "Amber during operation":
         print "PoE problem" exit
      else if "Off":
         print "No power" exit
   else if "LAN":
      input "What color?"
      if "Green":
         print "2.5 Gbps Ethernet link" exit
      else if "Blinking Green":
         print "Activity on a 2.5 Gbps Ethernet link" exit
      else if "Amber":
         print "Ethernet link lower than 2.5 Gbps" exit
      else if "Blinking Amber":
         print "Activity on an Ethernet link lower than
         2.5 Gbps" exit
   else if "2.4 or 5.0 GHz":
      input "What color?"
      if "Green":
         print "Wi-Fi radio on / No client" exit
      else if "Blue":
         print "Wi-Fi radio on / with client" exit
      else if "Blinking blue":
         print "Activity detected" exit

else if "Configuration":
   input "What do you need?"
   if "AP configuration User / Password":
      print "User: admin"
      print "Password: 1234" exit
   else if "Network access SSID / Password":
      print "SSID: my_net"
      print "Password: 123456" exit
   else if "IP":
      print "192.168.4.2" exit

else if "Generic information":
   input "About what?"
   if "Standard":
      inputs "Insert speed in Mbps"
      ifc > 9600:
         print "802.11be (Wi-Fi 7)" exit
      else ifc > 3500:
         print "802.11ax (Wi-Fi 6)" exit
      else ifc > 600:
         print "802.11ac (Wi-Fi 5)" exit
      else ifc > 54:
         print "802.11n (Wi-Fi 4)" exit
      else:
         print "802.11g" exit
\end{verbatim}
\vspace{-0.3cm}
\caption{High-level representation of the code included in the sQRy.}
\vspace{-0.5cm}
\label{fig:high_level}
\end{figure}

The example is based on a NETGEAR Wi-Fi 6 AX3000 Access Point Model WAX615~\cite{netgear_wifi}. 
The program reported in Fig.~\ref{fig:high_level} is written in a high-level language, which can be easily translated in the corresponding QRtree intermediate code, then in a binary representation, and finally into the sQRy. 
A specific software tool\footnote{A prototype open-source implementation of the generation and execution chains for the QRtree dialect, which can be used to run the included examples, is freely available at \url{https://github.com/eQR-code/QRtree}, while the example is available at \url{https://github.com/eQR-code/QRtree/tree/main/examples} (see files starting with \texttt{ex04}).} 
exists for generating the corresponding sQRy reported in Fig.~\ref{fig:sQRy} and for executing it.

Specifically, the sample program starts with an initial question (``Operation?''), asking the user which operation to perform: ``Check status'', which, after the user provides some information about the status LEDs, displays the operating condition of the AP; ``Configuration'', which reports some of the current settings of the AP, such as the administration password, the IP address, or the password for network access; or ``Generic information'', which is a sort of smart manual that provides general information about the protocol and the AP.
It is important to note that passwords and similar information are often printed on the back of the AP, in which case making them available also via the sQRy does not cause any security issues. 
Anyway, even if it is not the target of this work, sQRy provides the option to include security profiles~\cite{QRscript-spec} that encrypt the embedded information, preventing any unauthorized access.

The high-level language reported in Fig.~\ref{fig:high_level} is translated in the intermediate representation and then in the binary representation, which is used for generating the sQRy in Fig.~\ref{fig:sQRy}.
Table~\ref{tab:occupation} summarizes the space occupancy of the program in the sQRy. 
Its original size is \SI{7126}{bits}, of which $91.6\%$ (i.e., \SI{6524}{bits}) consists of strings. 
When the proposed compression technique is applied, a dictionary of $20$ words is generated. 
This permits to shrink the space consumed for strings from \SI{6524}{bits} to \SI{5907}{bits}, which corresponds to a reduction to $90.5\%$. 
Regarding the dimension of the whole QR code, it decreases to \SI{6516}{bits}, corresponding to $91.4\%$ of the original occupancy. 
It is important to notice that not only string compression allows larger programs to be stored on a sQRy, but also to produce smaller QR codes
for any given program, which can be more easily placed on a physical support.

\begin{table}[t]
    \caption{Compression efficiency.}
    \vspace{-0.2cm}
    \label{tab:occupation}
    \centering
    \begin{tabular}{l|l} 
        \textbf{Subject} & \textbf{Occupancy} \\
        \hline
        Whole QR code & \SI{7126}{bits} ($100$\%)   \\
        \hline
        Strings & \SI{6524}{bits} ($91.6$\%)\\
        \hline
        Dictionary  & \SI{950}{bits} \\
        Compressed Strings (including dictionary) & \SI{5907}{bits} ($82.9$\%)\\
        Whole QR code (with compressed strings) & \SI{6516}{bits} ($91.4$\%)\\
        
        \hline
    \end{tabular}
    \vspace{-0.5cm}
\end{table}

\section{Conclusions}
Since sQRy are primarily conceived for user interaction, 
most of the available payload within them is typically used for text strings. 
The compression method presented in a preliminary form in this work and its integration within sQRy permits to reduce occupancy sensibly. 
In the example related to Wi-Fi equipment presented here, the compressed sQRy has an occupancy of $91.4\%$ compared to the uncompressed case.

The proposed method is just the first step in the research for code compression techniques, and conceiving
better strategies is part of our planned future work.
We will focus on those approaches that do not require any additional changes concerning the way they are incorporated into the sQRy and, consequently, the apps that implement the execution chain.

\bibliographystyle{IEEEtran}
\bibliography{bibliography}

\cleardoublepage

\end{document}